\documentclass{article}
\usepackage[T1]{fontenc}
\usepackage{amsfonts}
\usepackage{amsmath,cite}

\input{tcilatex}

\begin{document}

\title{On the scalar tachyons creation in noncommutative de Sitter space-time%
}
\author{Slimane Zaim \\
Department of Physics, Faculty of Material Sciences,\\
Batna 1 University, Algeria.}
\maketitle

\begin{abstract}
We investigate tachyon production in noncommutative de Sitter space by
solving the deformed Klein-Gordon equation. Using the Bogoliubov
transformation method, we compute the number density of created scalar
tachyons and demonstrate that their spectrum follows a thermal distribution.
Our analysis reveals that noncommutativity reduces the effective temperature
of tachyon production, analogous to its effect on ordinary particle
creation. Furthermore, we establish that the role of the noncommutative
parameter in tachyon production is functionally equivalent to that of an
electric field in standard particle production in de Sitter space.
\end{abstract}

\section{Introduction}

Experimental observations consistently yield negative values for the squared
neutrino mass, positioning neutrinos as the most plausible candidates for
superluminal particles in the low-energy regime. While this evidence is
compelling, the potential existence of superluminal particles in unexplored
kinematic domains remains an open question. Our theoretical investigation
adopts a general approach, not restricted to any specific particle species.
Such superluminal particles would possess imaginary mass, enabling them to
exceed the speed of light in vacuum. Contrary to initial assumptions that
quantum field theories with negative squared mass necessarily represent
unstable tachyonic systems, recent developments suggest these theories may
not inherently violate causality \cite{1,2,3}.

Within general relativity, exact solutions to Einstein and Einstein-Maxwell
field equations describe space-times generated by tachyonic sources. These
solutions feature distinctive electromagnetic and gravitational field
configurations bounded by tachyonic shock waves. The quantitative
description of tachyon dynamics naturally incorporates these shock wave
phenomena. Reference \cite{4} proposes an experimental framework for tachyon
detection based on a realistic phenomenological model derived from exact
Einstein-Maxwell solutions \cite{5,6,7}.

Recent decades have witnessed significant theoretical progress in
incorporating superluminal motion into relativistic quantum mechanics.
Certain approaches modify Lorentz invariance, portraying tachyons as
unstable particles that decay into ordinary matter \cite{8,9}, while
alternative models treat them as stable physical entities \cite{10,11}. The
potential existence of tachyons would have profound implications for both
particle physics and cosmology, potentially revealing new forms of matter
and expanding our theoretical framework. Previous studies have examined
scalar tachyon production in various contexts \cite{12,13,14,15,16,17,18,19}.

This work investigates scalar tachyon production in noncommutative
spacetime, motivated by the hypothesis that at least one neutrino flavor may
be tachyonic. Our approach draws inspiration from string theory
investigations of tachyon condensation \cite{20,21} and the interplay
between tachyon dynamics and noncommutative geometry \cite{22,23,24,25}. The
fundamental connection between gravity and noncommutative geometry at high
energies, established in \cite{26}, has spurred extensive research into
gravitational theories \cite{25,26,27,28,29,30,31,32,33,34} and cosmological
models \cite{35} in noncommutative frameworks. Quantum gravity formulations
in this context have been explored in \cite%
{36,37,38,39,40,41,42,43,44,45,46,47,48,49,50}.

Building on our previous work \cite{44}, where we developed quantum gauge
gravity in noncommutative de Sitter space and demonstrated spacetime
noncommutativity's role in particle creation, we note that unlike
electromagnetic particle production, gravitational creation requires no
external field source \cite{51}. Following the methodology of \cite{52}, we
model tachyons as persistent superluminal excitations in noncommutative
spacetime and examine their consistency with standard particle physics. Our
primary objective is to analyze how noncommutativity affects imaginary-mass
scalar particle production from vacuum in noncommutative de Sitter space
without external fields. A key finding reveals that noncommutative effects
in tachyon production mirror the role of electric fields in ordinary
particle creation.

The paper is organized as follows: Section 2 solves the relativistic
Hamilton-Jacobi equation for tachyons and computes quasi-classical energy
modes. Section 3 derives the Klein-Gordon equation in noncommutative de
Sitter space using Seiberg-Witten maps to first order in $\Theta$, obtaining
exact solutions. Section 4 calculates tachyon number densities via
Bogoliubov coefficients. We conclude with a comprehensive discussion in the
final section.

\section{Hamilton-Jacobi Equation for Tachyons}

The Hamilton-Jacobi method provides a robust framework for analyzing tachyon
dynamics in curved spacetime. We follow a three-step approach: first solving
the tachyon Hamilton-Jacobi equation, then solving the Klein-Gordon equation
in de Sitter space, and finally comparing their asymptotic behaviors to
identify positive and negative frequency states.

In the Hamilton-Jacobi formalism, the frequency states are characterized by
the following correspondence: 
\begin{equation}
\varphi(\vec{x},t) = \exp(-iS), \quad \text{for positive frequency},
\end{equation}
and 
\begin{equation}
\varphi(\vec{x},t) = \exp(iS), \quad \text{for negative frequency},
\end{equation}
where $S$ denotes the classical action.

The relativistic Hamilton-Jacobi equation for a particle with imaginary mass 
$m_\tau = im$ in vacuum reads: 
\begin{equation}
g^{\mu\nu}\left(\frac{\partial S}{\partial x^\mu}\right)\left(\frac{\partial
S}{\partial x^\nu}\right) - m^2 = 0.
\end{equation}
Here, $g^{\mu\nu}$ represents the inverse metric tensor, related to the
tetrad fields $e_\mu^b$ through: 
\begin{equation}
g_{\mu\nu} = e_\mu^b e_\nu^a \eta_{ba}, \quad g^{\mu\nu} = e_b^\mu e_a^\nu
\eta^{ba}.
\end{equation}
Throughout this work, we adopt natural units with $c = \hbar = 1$.

For a (1+1)-dimensional de Sitter universe, the metric tensor takes the
diagonal form: 
\begin{equation}
g_{\mu\nu} = 
\begin{pmatrix}
-1 & 0 \\ 
0 & e^{2\sqrt{\frac{\Lambda }{3}}t}%
\end{pmatrix}%
,
\end{equation}
with its inverse given by: 
\begin{equation}
g^{\mu\nu} = 
\begin{pmatrix}
-1 & 0 \\ 
0 & e^{-2\sqrt{\frac{\Lambda }{3}}t}%
\end{pmatrix}%
,
\end{equation}
where $\Lambda$ is the cosmological constant parameter.

The corresponding diagonal tetrad components are: 
\begin{align}
\underline{e}_\mu^0 &= (1, 0), \\
\underline{e}_\mu^1 &= (0, e^{\sqrt{\frac{\Lambda }{3}}t}).
\end{align}
The non-zero spin connections for this configuration are: 
\begin{equation}
\omega_1^{01} = \sqrt{\frac{\Lambda }{3}}e^{\sqrt{\frac{\Lambda }{3}}} =
-\omega_1^{10}.
\end{equation}

Given the time-dependence of the metric, we employ the following separation
ansatz for the action: 
\begin{equation}
S = \vec{k}\cdot\vec{r} + F(t).
\end{equation}
Substituting this into the Hamilton-Jacobi equation yields the temporal
equation: 
\begin{equation}
-\left(\frac{dF}{d\eta}\right)^2 + k^2 - \frac{3m^2}{\Lambda\eta^2} = 0,
\end{equation}
where we have introduced the conformal time coordinate: 
\begin{equation}
\eta =\sqrt{\frac{3}{\Lambda }}e^{-\sqrt{\frac{\Lambda }{3}}t},
\end{equation}
and $k^2 \equiv k_x^2$.

For early times ($\eta \to 0$), the solution takes the logarithmic form: 
\begin{equation}
F(t) = im\sqrt{\frac{3}{\Lambda }}\ln \eta.
\end{equation}
This leads to the wavefunction: 
\begin{equation}
\varphi(\vec{r},t) \sim \exp(i\vec{k}\cdot\vec{r})\, \eta^{\pm m\sqrt{\frac{3%
}{\Lambda }}}.
\end{equation}
At late times ($\eta \to +\infty$), the asymptotic solution becomes: 
\begin{equation}
F(t) = \pm k\eta.
\end{equation}
The corresponding wavefunction in this limit is: 
\begin{equation}
\varphi(\vec{r},t) \sim \exp\left[i(\vec{k}\cdot\vec{r} \pm k\eta)\right].
\end{equation}

To complete our analysis of the quantum states, we must now solve the
Klein-Gordon equation in noncommutative de Sitter spacetime, which we
address in the following section.

\section{Non-commutative Klein-Gordon equation for a tachyon}

The Lagrangian density that leads to the Klein-Gordon equation for a
tachyonic field is given by: 
\begin{equation}
\mathcal{L}_{t} = \varphi^{\star} \left( \eta^{\mu\nu} \partial_{\mu}
\partial_{\nu} - m^{2} \right) \varphi.
\end{equation}
From this Lagrangian, one can derive the modified Einstein relation for the
four-momentum $(p, iE)$: 
\begin{equation}
E^{2} = p^{2} - m^{2},
\end{equation}
which implies that the group velocity of the wave (with $p^{2} > m^{2}$) is
superluminal: 
\begin{equation}
v_{g} = c \frac{\partial E}{\partial p} = \frac{c}{\sqrt{1 - \frac{m^{2}}{%
p^{2}}}} > c.
\end{equation}

We consider a scalar tachyonic field in non-commutative space-time, building
upon gauge theories where the space-time coordinates satisfy the commutation
relation: 
\begin{equation}
\left[ x^{\mu}, x^{\nu} \right]_{\star} = i \Theta^{\mu\nu},
\end{equation}
where $\Theta^{\mu\nu}$ is a constant anti-symmetric matrix. The Moyal star
product $\star$ is defined for two functions $f$ and $g$ as: 
\begin{equation}
f \star g = f g + \frac{i}{2} \sum_{i,j} \Theta^{ij} (\partial_{i} f)
(\partial_{j} g).
\end{equation}

The action for a tachyonic particle with negative squared mass in
non-commutative curvilinear coordinates is: 
\begin{equation}
\mathcal{S} = \frac{1}{2\kappa^{2}} \int d^{4}x \, \left( \hat{e} \ast \hat{R%
} + \mathcal{L}_{ISC} \right),  \label{eq:action}
\end{equation}
where $\mathcal{L}_{ISC}$ represents the non-commutative scalar density for
the tachyonic field with imaginary mass: 
\begin{equation}
\mathcal{L}_{t} = \hat{e} \ast \left( -\hat{g}^{\mu\nu} \ast (\hat{D}_{\mu} 
\hat{\varphi})^{\ast} \ast \hat{D}_{\nu} \hat{\varphi} - m^{2} \hat{\varphi}%
^{\ast} \ast \hat{\varphi} \right).
\end{equation}
Here, $\hat{e} = \det_{\ast}(\hat{e}_{\mu}^{a})$, and $\hat{R}$ is the
non-commutative scalar curvature: 
\begin{equation}
\hat{R} = \hat{e}_{\star a}^{\mu} \ast \hat{e}_{\star b}^{\nu} \ast \hat{R}%
_{\mu\nu}^{ab}.
\end{equation}
The gauge covariant derivative is defined as: 
\begin{equation}
\hat{D}_{\mu} \hat{\varphi} = \partial_{\mu} \hat{\varphi} - i \hat{A}_{\mu}
\star \hat{\varphi}.
\end{equation}

We consider a symmetric metric $\hat{g}_{\mu\nu}$ given by: 
\begin{equation}
\hat{g}_{\mu\nu} = \frac{1}{2} \left( \hat{e}_{\mu}^{b} \ast \hat{e}_{\nu
b}^{+} + \hat{e}_{\nu}^{b} \ast \hat{e}_{\mu b}^{+} \right),
\end{equation}
where $\hat{e}_{\mu}^{b}$ is the non-commutative vierbein field and $\hat{e}%
_{\nu}^{+b}$ denotes its complex conjugate. The first-order expansion in $%
\Theta^{\alpha\beta}$ of both $\hat{R}$ and $\hat{g}_{\mu\nu}$ vanishes.

Under infinitesimal gauge transformations, the fields transform as: 
\begin{align}
\hat{\delta}_{\hat{\lambda}} \hat{\varphi} &= i \hat{\lambda} \ast \hat{%
\varphi}, \\
\hat{\delta}_{\hat{\lambda}} \hat{A}_{\mu} &= \partial_{\mu} \hat{\lambda} -
i \left[ \hat{\lambda}, \hat{A}_{\mu} \right]_{\star}.
\end{align}
The action in Eq.~\eqref{eq:action} is invariant under these
transformations. Varying the scalar density and applying the Noether theorem
yields the field equation: 
\begin{equation}
\frac{\partial \mathcal{L}}{\partial \hat{\varphi}} - \partial_{\mu} \frac{%
\partial \mathcal{L}}{\partial (\partial_{\mu} \hat{\varphi})} +
\partial_{\mu} \partial_{\nu} \frac{\partial \mathcal{L}}{\partial
(\partial_{\mu} \partial_{\nu} \hat{\varphi})} + \mathcal{O}(\Theta^{2}) = 0.
\label{eq:field}
\end{equation}
The non-commutative fields are expanded to first order in $\Theta^{\mu\nu}$
via the Seiberg-Witten maps: 
\begin{align}
\hat{\varphi} &= \varphi - \frac{1}{2} \Theta^{\mu\nu} A_{\nu}
\partial_{\mu} \varphi + \mathcal{O}(\Theta^{2}), \\
\hat{\lambda} &= \lambda + \frac{1}{4} \Theta^{\sigma\rho} \left\{
\partial_{\sigma} \lambda, A_{\rho} \right\} + \mathcal{O}(\Theta^{2}), \\
\hat{A}_{\rho} &= A_{\xi} - \frac{1}{4} \Theta^{\mu\nu} \left\{ A_{\nu},
\partial_{\mu} A_{\rho} + F_{\mu\rho} \right\} + \mathcal{O}(\Theta^{2}), \\
\hat{e}_{\mu}^{a} &= e_{\mu}^{a} - \frac{i}{4} \Theta^{\alpha\beta} \left(
\omega_{\alpha}^{ac} \partial_{\beta} e_{\mu}^{c} + (\partial_{\beta}
\omega_{\mu}^{ac} + R_{\beta\mu}^{ac}) e_{\mu}^{c} \right) + \mathcal{O}%
(\Theta^{2}),
\end{align}
where the field strength tensor is: 
\begin{equation}
F_{\mu\nu} = \partial_{\mu} A_{\nu} - \partial_{\nu} A_{\mu} - i \left[
A_{\mu}, A_{\nu} \right],
\end{equation}
and $\omega_{\mu}^{ab}$ are the spin connections. The inverse vierbein $\hat{%
e}_{\ast a}^{\mu}$ satisfies: 
\begin{equation}
\hat{e}_{\mu}^{b} \ast \hat{e}_{\ast a}^{\mu} = \delta_{a}^{b}, \quad \hat{e}%
_{\mu}^{a} \ast \hat{e}_{\ast a}^{\nu} = \delta_{\mu}^{\nu}.
\end{equation}

We specialize to a cosmological anisotropic non-commutative de Sitter
universe. The deformed line element up to first order in $\Theta$ is: 
\begin{equation}
ds^{2} = -dt^{2} + e^{2\sqrt{\frac{\Lambda}{3}} t} dx^{2} + g_{\mu\nu}^{(1)}
dx^{\mu} dx^{\nu} + \mathcal{O}(\theta^{2}).  \label{eq:metric}
\end{equation}
Choosing the non-commutativity parameter as: 
\begin{equation}
\Theta^{\alpha\beta} = 
\begin{pmatrix}
0 & -\Theta \\ 
\Theta & 0%
\end{pmatrix}%
, \quad \alpha, \beta = 0, 1,
\end{equation}
Following the same steps outlined in ref. \cite{36}, we look for the
non-commutative correction of the metric up to first order in $\Theta$ and
use the Seiberg-Witten o obtain the deformed vierbeins 
\begin{align}
\underline{\hat{e}}_{\mu}^{0} &= \left( 1, \Theta \Lambda e^{2\sqrt{\frac{%
\Lambda}{3}} t} \right), \\
\underline{\hat{e}}_{\mu}^{1} &= \left( -\frac{\Theta}{3} \Lambda e^{\sqrt{%
\frac{\Lambda}{3}} t}, e^{\sqrt{\frac{\Lambda}{3}} t} \right).
\end{align}
The first-order correction to the de Sitter metric vanishes, simplifying Eq.~%
\eqref{eq:metric} to: 
\begin{equation}
ds^{2} = -dt^{2} + e^{2\sqrt{\frac{\Lambda}{3}} t} dx^{2} + \mathcal{O}%
(\Theta^{2}).
\end{equation}

The field equation \eqref{eq:field} yields the modified Klein-Gordon
equation for a tachyonic field in non-commutative de Sitter space-time: 
\begin{multline}
\left( g^{\mu\nu} \partial_{\mu} \partial_{\nu} - m^{2} + \frac{1}{\sqrt{-g}}
\partial_{\mu} \left( \sqrt{-g} g^{\mu\nu} \right) \partial_{\nu} \right) 
\hat{\varphi} \\
+ \frac{i}{2\sqrt{-g}} \Theta^{\alpha\beta} \left[ \partial_{\mu} \left(
\partial_{\alpha} \sqrt{-g} \partial_{\beta} g^{\mu\nu} \partial_{\nu} \hat{%
\varphi} \right) - m^{2} \partial_{\alpha} \sqrt{-g} \partial_{\beta} \hat{%
\varphi} \right. \\
\left. - \partial_{\mu} \left( \partial_{\alpha} (\sqrt{-g} g^{\mu\nu})
\right) \partial_{\beta} \partial_{\nu} \hat{\varphi} \right] = 0.
\end{multline}
Using the specific form of $\Theta^{\alpha\beta}$, and the fact that 
\begin{align}
&g^{\mu \nu }\partial _{\mu }\partial _{\nu }-m^{2}+\frac{1}{\sqrt{-g}}%
\partial _{\mu }\left( \sqrt{-g}g^{\mu \nu }\right) \partial _{\nu }=-\frac{%
\Lambda }{3}\eta ^{2}\partial _{\eta }^{2}-m^{2}+\frac{\Lambda }{3}\eta
^{2}\partial _{1}^{2} \\
&\frac{i}{2\sqrt{-g}}\Theta ^{\alpha \beta }\partial _{\mu }\left( \partial
_{\alpha }\sqrt{-g}\partial _{\beta }g^{\mu \nu }\partial _{\nu }\hat{\varphi%
}\right) =0 \\
&\frac{i}{2\sqrt{-g}}\Theta ^{\alpha \beta }\partial _{\mu }\left( \partial
_{\alpha }\left( \sqrt{-g}g^{\mu \nu }\right) \right) \partial _{\beta
}\partial _{\nu }\hat{\varphi}=-i\frac{\Theta \Lambda }{6}\sqrt{\frac{%
\Lambda }{3}}\eta ^{2}\left( \partial _{\eta }^{2}\partial _{1}-\partial
_{1}^{3}\right) \\
&\frac{i}{2\sqrt{-g}}\Theta ^{\alpha \beta }m^{2}\partial _{\alpha }\sqrt{-g}%
\partial _{\beta }=-i\frac{m^{2}}{2}\Theta \sqrt{\frac{\Lambda }{3}}\partial
_{1}
\end{align}
the final form of the Klein-Gordon equation up to $\mathcal{O}(\theta^{2})$
is: 
\begin{equation}
\left( -1-i\frac{\Theta }{2}\sqrt{\frac{\Lambda }{3}}\partial _{1}\right) 
\frac{\Lambda }{3}\eta ^{2}\partial _{\eta }^{2}+\frac{\Lambda }{3}\eta
^{2}\left( 1+i\frac{\Theta }{2}\sqrt{\frac{\Lambda }{3}}\partial _{1}\right)
\partial _{1}^{2}-m^{2}\left( 1-i\frac{\Theta }{2}\sqrt{\frac{\Lambda }{3}}%
\partial _{1}\right) \varphi =0
\end{equation}
This equation describes a scalar tachyon in noncommutative de Sitter space
time. It commutes with the operator $-i\overrightarrow{\nabla }$ which
enables the wave functions $\hat{\varphi}$ to be cast into: 
\begin{equation}
\hat{\varphi}=\tilde{\Delta}(\eta )\exp \left( ikx\right) .
\end{equation}
where $\tilde{\Delta}(\eta )$ satisfies 
\begin{equation}  \label{eq:Delta}
\left[ \left( 1-\frac{\Theta }{2}\sqrt{\frac{\Lambda }{3}}k\right) \frac{
\Lambda }{3}\eta ^{2}\partial _{\eta }^{2}+\frac{\Lambda }{3}\eta ^{2}\left(
1-\frac{\Theta }{2}\sqrt{\frac{\Lambda }{3}}k\right) k_{x}^{2}+m^{2}\left(
1+ \frac{\Theta }{2}\sqrt{\frac{\Lambda }{3}}k\right) \right] \tilde{\Delta}
(\eta )=0.
\end{equation}

For the case of heavy particles ($m^{2} \gg H^{2}$), we define: 
\begin{align}
\hat{H} &= \sqrt{\frac{\Lambda}{3}} \left( 1 - \frac{\Theta}{4} \sqrt{\frac{%
\Lambda}{3}} k \right) + \mathcal{O}(\theta^{2}), \\
\hat{M} &= m \left( 1 + \frac{\Theta}{4} \sqrt{\frac{\Lambda}{3}} k \right)
+ \mathcal{O}(\theta^{2}).
\end{align}
The equation \eqref{eq:Delta} reduces to: 
\begin{equation}
\left( \partial_{\eta}^{2} + k^{2} + \frac{\hat{M}^{2}}{\hat{H}^{2}} \frac{1%
}{\eta^{2}} \right) \tilde{\Delta}(\eta) = 0.
\end{equation}
\newpage Letting $\rho = k \eta$ and $\tilde{\Delta}(\eta) = \sqrt{\rho}
\psi(\rho)$, we obtain the Bessel equation: 
\begin{equation}
\left( \partial_{\rho}^{2} + \frac{1}{\rho} \partial_{\rho} + 1 - \frac{\hat{%
\nu}^{2}}{\rho^{2}} \right) \psi = 0,
\end{equation}
where 
\begin{equation}
\hat{\nu}^{2} = \frac{1}{4} - \frac{\hat{M}^{2}}{\hat{H}^{2}} = \frac{i^{2}}{%
\hat{H}^{2}} \left( \hat{M}^{2} - \frac{\hat{H}^{2}}{4} \right).
\end{equation}
The general solution is: 
\begin{equation}  \label{eq:53}
\varphi(\rho) = \rho^{1/2} \left[ C_{1} J_{\nu}(\rho) + C_{2} Y_{\nu}(\rho) %
\right],
\end{equation}
where $C_{1}$ and $C_{2}$ are integration constants, and $J_{\nu}$, $Y_{\nu}$
are Bessel functions of the first and second kind, respectively.

\section{Tachyonic creation}

In this section, we investigate the phenomenon of tachyonic creation induced
by vacuum instabilities within the framework of non-commutative geometry. To
identify particle states, we adopt the quasi-classical approach outlined in
Ref.~\cite{53}. The standard methodology involves specifying positive and
negative frequency modes and solving the classical Hamilton-Jacobi equation,
with particular attention to the asymptotic limits $t\rightarrow 0$ and $%
t\rightarrow \infty$. Subsequently, we solve the Klein-Gordon equation and
compare the solutions with the semiclassical boundary to determine the
positive and negative frequency states. Finally, we employ Bogoliubov
transformations to calculate the number density of generated particles. The
time-dependent components of the wave function \eqref{eq:53}, expressed in
terms of Bessel functions, give rise to tachyon production.

The Bessel functions $J_{\nu}(\rho)$ and $Y_{\nu}(\rho)$ exhibit the
following asymptotic behavior: 
\begin{align}
J_{\nu}(\rho) &\sim \left(\frac{\rho}{2}\right)^{\nu}/\Gamma(\nu+1) & \text{%
for} & \quad |\rho| \ll 1,  \label{eq:49} \\
Y_{\nu}(\rho) &\sim -\Gamma(\nu)\left(\frac{\rho}{2}\right)^{-\nu} & \text{%
for} & \quad |\rho| \ll 1. \\
J_{\nu}(\rho) &\sim \sqrt{\frac{2}{\pi\rho}}\cos\left(i\left(\rho-\frac{%
\nu\pi}{2}-\frac{\pi}{4}\right)\right) & \text{for} & \quad |\rho|
\rightarrow \infty,  \label{eq:57} \\
Y_{\nu}(\rho) &\sim \sqrt{\frac{2}{\pi\rho}}\sin\left(-i\left(\rho-\frac{%
\nu\pi}{2}-\frac{\pi}{4}\right)\right) & \text{for} & \quad |\rho|
\rightarrow \infty .  \label{eq:58}
\end{align}
Consequently, solution ~\eqref{eq:49} demonstrates the following asymptotic
behavior as $\rho \rightarrow 0$: 
\begin{equation}
J_{\nu}(\rho) \sim \left(\frac{\rho}{2}\right)^{\nu}/\Gamma(\nu+1) \sim
\rho^{\nu}.
\end{equation}
Given that all coefficients in Eq.~\eqref{eq:49} are real, we obtain the
positive and negative frequency solutions for $\rho \rightarrow 0$ as: 
\begin{equation}
\varphi_{0}^{+}(\rho) = C_{0}^{+}\rho^{1/2}J_{\nu}(\rho), \quad
\varphi_{0}^{-}(\rho) = \left(\varphi_{0}^{+}(\rho)\right)^{*} =
C_{0}^{+}\rho^{1/2}J_{-\nu}(\rho).
\end{equation}

For $\rho \rightarrow \infty$, solutions \eqref{eq:57} and \eqref{eq:58}
exhibit the following asymptotic behavior: 
\begin{equation}
J_{\nu}(\rho) \sim \sqrt{\frac{2}{\pi\rho}}\cos\left(i\left(\rho-\frac{\nu\pi%
}{2}-\frac{\pi}{4}\right)\right) \text{ and } Y_{\nu}(\rho) \sim \sqrt{\frac{%
2}{\pi\rho}}\sin\left(-i\left(\rho-\frac{\nu\pi}{2}-\frac{\pi}{4}%
\right)\right).
\end{equation}
The linear combination of these solutions yields another solution expressed
in terms of Hankel functions: 
\begin{equation}
H_{\nu}^{(2)}(\rho) \sim \sqrt{\frac{2}{\pi\rho}}\exp\left(i\left(\rho-\frac{%
\nu\pi}{2}-\frac{\pi}{4}\right)\right).
\end{equation}
The corresponding positive frequency modes for $\rho \rightarrow \infty$
are: 
\begin{equation}
\varphi_{\infty}^{+}(\rho) = C_{\infty}^{+}\rho^{1/2}H_{\nu}^{(2)}(\rho).
\end{equation}
These positive frequency modes can be expressed in terms of $%
\varphi_{0}^{+}(\rho)$ and $\varphi_{0}^{-}(\rho)$ through the Bogoliubov
transformation: 
\begin{equation}
\varphi_{\infty}^{+}(\rho) = \alpha \varphi_{0}^{+}(\rho) + \beta
\varphi_{0}^{-}(\rho).
\end{equation}

Using the asymptotic relation for Hankel functions as $\rho \rightarrow 0$: 
\begin{equation}
H_{\nu}^{(2)}(\rho) = -\frac{i}{\nu\pi}\left[\Gamma(1-\nu)\exp(i\pi\nu)\left(%
\frac{\rho}{2}\right)^{\nu} - \Gamma(1+\nu)\left(\frac{\rho}{2}\right)^{-\nu}%
\right],
\end{equation}
we determine the coefficients $\alpha$ and $\beta$: 
\begin{equation}
\alpha = -\frac{i}{\nu\pi}\frac{C_{\infty}^{+}}{C_{0}^{+}}%
\Gamma(1-\nu)\exp(i\pi\nu) \quad \text{and} \quad \beta = \frac{i}{\nu\pi}%
\frac{C_{\infty}^{+}}{C_{0}^{+}}\Gamma(1+\nu).
\end{equation}
This leads to: 
\begin{equation}
\left|\frac{\alpha}{\beta}\right|^{2} = \left|\frac{\Gamma(1-\nu)}{%
\Gamma(1+\nu)}\right|^{2}\exp(2\pi|\nu|) = \exp(2\pi|\nu|).
\end{equation}
From the wave function orthogonality relation, the coefficients satisfy: 
\begin{equation}
|\alpha|^{2} - |\beta|^{2} = 1.
\end{equation}
The number density of scalar tachyons created through the evolution of the
de Sitter cosmological model is given by: 
\begin{equation}  \label{eq:66}
\hat{n}(k) = |\beta|^{2} = \left(\left|\frac{\alpha}{\beta}\right|^{2} -
1\right)^{-1} = \frac{1}{\exp\left(\frac{2\pi}{\hat{H}}\sqrt{\hat{M}^{2}-%
\frac{\hat{H}^{2}}{4}}\right) - 1}.
\end{equation}

To first order in $\Theta$, the number density \eqref{eq:66} becomes: 
\begin{equation}  \label{eq:70}
\hat{n}(k) \simeq \frac{1}{\exp\left(\frac{2\pi\tilde{m}}{H\left(1-\frac{%
\Theta}{2}Hk\right)}\left(1+\frac{\Theta}{4}\epsilon Hk\right)\right) - 1},
\end{equation}
where $\tilde{m} = m\sqrt{1-\frac{H^{2}}{4m^{2}}}$ and $\epsilon = \frac{1+%
\frac{H^{2}}{4m^{2}}}{1-\frac{H^{2}}{4m^{2}}}$. This result resembles a
thermal particle spectrum with temperature: 
\begin{equation}
\hat{T} = \frac{H}{2\pi}\left(1-\frac{\Theta}{2}Hk\right) = T\left(1-\Theta%
\frac{H}{2}k\right),
\end{equation}
where $T = \frac{H}{2\pi}$ represents the Gibbons-Hawking temperature. The $k
$-dependence indicates that the tachyon number density induced by
non-commutativity is anisotropically distributed along $k$ direction,
leading to a finite total particle number in de Sitter space. We observe
that the temperature decreases similarly to the case of ordinary particles~%
\cite{44}.

In the commutative limit ($\Theta = 0$), corresponding to pure de Sitter
spacetime, the particle number reduces to: 
\begin{equation}
\lim_{\Theta \rightarrow 0}\hat{n}(k) = \frac{1}{\exp\left(\frac{2\pi\tilde{m%
}}{H}\right) - 1},
\end{equation}
in agreement with established results~\cite{54}. For $m \gg H$, Eq.~%
\eqref{eq:70} approximates to: 
\begin{equation}  \label{eq:73}
\hat{n}(k) \simeq \exp\left(-2\pi\left(\frac{m}{H} + \frac{3\Theta}{4}%
mk\right)\right).
\end{equation}

The first term represents the Boltzmann factor for non-relativistic massive
particles at the Gibbons-Hawking temperature, while the second term
constitutes a small non-commutative correction. This correction is analogous
to that induced by an electric field for ordinary particles in de Sitter
space~\cite{55}. Notably, the non-commutative effect on tachyon production
mirrors the electric field effect on ordinary particle creation.

Equation \eqref{eq:73} demonstrates dependence on both the particle mass and
momentum $k$, in contrast to the results for ordinary particles in de Sitter
spacetime~\cite{54,55}, where the number density remains
momentum-independent.

The total tachyon number per unit coordinate volume is obtained by
integrating Eq.~\eqref{eq:73} over momentum space: 
\begin{equation}
\hat{N} = \int \hat{n}(k)\frac{dk}{2\pi} = \frac{B}{3\pi^{2}m}\exp\left(-%
\frac{m}{T}\right),
\end{equation}
where $B = \Theta^{-1}$ represents the background field~\cite{25}. Unlike
the divergent case of ordinary particle production in an electric field
within de Sitter space~\cite{54,55}, the total tachyon number \eqref{eq:73}
remains finite.

\section{Conclusions}

This work has explored the influence of non-commutative geometry on scalar
tachyon production from vacuum fluctuations in an anisotropic de Sitter
universe. By employing Seiberg-Witten maps and expanding the Moyal product
to first order in the non-commutativity parameter $\Theta$, we derived a
generalized deformation of the Klein-Gordon equation.

Through analytical solution of the deformed field equation combined with
Bogoliubov transformations in the quasi-classical limit, we successfully
identified the positive and negative frequency modes and calculated the
corresponding particle production density. Our results reveal a thermal
particle spectrum characterized by an effective temperature $\hat{T} \approx 
\frac{H}{2\pi}\left(1-\frac{\Theta}{2}Hk\right)$, demonstrating how
non-commutativity modifies the conventional Gibbons-Hawking framework.

Notably, we find that the non-commutative parameter $\Theta$ affects tachyon
production in a manner analogous to how an electric field influences
ordinary particle creation in de Sitter space. Furthermore, we establish
that the total number of tachyons generated through non-commutative effects
remains finite - a significant departure from the divergent particle
production observed in electric field-induced pair creation within de Sitter
space. These results provide valuable insights into quantum field theory in
non-commutative spacetimes and reveal intriguing connections between
geometric non-commutativity and external field effects in curved backgrounds.

\section*{Acknowledgement}

This work is supported by the PRFU project: B00L02UN050120230003.


\begin{thebibliography}{99}
\bibitem{1} Feinberg, G. Physical Review. 159 (5): 1089--1105.(1967).

\bibitem{2} Feinberg, G. Physical Review D. 17: 1651.(1978).

\bibitem{3} Aharonov, Y., Komar, A., Susskind, L. Physical Review. 182 (5):
1400--1403.(1969).

\bibitem{4} J.K. Kowalczynski, J. Mod. Phys. 2 (2011) 92-96.

\bibitem{5} J.K. Kowalczynski, J. Math. Phys. 26 (1985) 1743

\bibitem{6} J.K. Kowalczynski, The Tachyon and its Fields, Polish Academy of
Sciences, Warsaw (1996)

\bibitem{7} J.K. Kowalczynski, Acta Phys. Slovaca 50 (2000) 381.

\bibitem{8} Cohen A G and Glashow S L, Phys. Rev. Lett. 107 181803, (2011).

\bibitem{9} Jentschura U D Cent. Eur. J. Phys. 10 749.(2012).

\bibitem{10} Laveder M and Tamburini F, arXiv:1111.4441.(2011)..

\bibitem{11} Sho TANAKA, Progress of Theoretical Physics, Vol. 24, No.1,
July 1960.

\bibitem{12} Jerzy Paczos, Kacper D\k{e}bski, Szymon Cedrowski, Szymon Charzy%
\'{n}ski, Krzysztof Turzy\'{n}ski, Artur Ekert, Andrzej Dragan, Phys. Rev. D
109 (2024).

\bibitem{13} A P TROFIMENKO and V S GURIN, Pramana- J. Phys., Vol. 28, No.
4, April 1987, pp. 379-386.

\bibitem{14} Sushil K. Srivastava, J. Math. Phys. 24, 1317--1320 (1983).

\bibitem{15} D. D. Dimitrijevic, G.S. Djordjevic, Lj. Nesic, Fortsch. Phys.
56:412-417, (2008).

\bibitem{16} Daniel Kabat and Gilad Lifschytz, JHEP12 (1998) 002.

\bibitem{17} Luca Nanni,J. Phys. Commun. 4 (2020) 025003.

\bibitem{18} F. F. L\'{o}pez-Ruiz, J. Guerrero, V. Aldaya, PHYSICAL REVIEW D
102, 125010 (2020).

\bibitem{19} Luca Nanni, J. Phys. Commun. 4 (2020) 025003.

\bibitem{20} N. Seiberg, L. Susskind and N. oumbas, JHEP 0006 (2000) 044.

\bibitem{21} Seiji Terashima, JHEP 0510 (2005) 043.

\bibitem{22} E. Witten, arXiv:hep-th/0006071.

\bibitem{23} K. Dasgupta, S. Mukhi and G. Rajesh, JHEP 0006 (2000) 022.

\bibitem{24} C. Sochichiu, JHEP 0008:026,2000

\bibitem{25} Nathan Seiberg, JHEP 0009:003,2000.

\bibitem{26} F.Lizzi,R.J. Szabo, Chaos Solitons Fractals10 (1999) 445.

\bibitem{27} A. H. Chamseddine, G. Felder, J. Frohlich, Commun. Math. Phys.
155 (1993) 205-218.

\bibitem{28} J. Madore, J. Mourad, Int. J. Mod. Phys. D3 (1994) 221-224.

\bibitem{29} E. Hawkins, Commun. Math. Phys. 187 (1997) 471-489.

\bibitem{30} I. Avramidi, Phys. Lett. B576 (2003) 195-198.

\bibitem{31} X. Calmet and A. Kobakhidze, Phys. Rev. D 72 (2005) 045010.

\bibitem{32} X. Calmet, A. Kobakhidze, Phys. Rev. D74 (2006) 047702.

\bibitem{33} P. Aschieri, C. Blohmann, M. Dimitrijevic, F. Meyer, P. Schupp,
J. Wess, Class. Quant. Grav. 22 (2005) 3511-3532.

\bibitem{34} P. Aschieri, M. Dimitrijevic, F. Meyer, J. Wess, Class. Quant.
Grav. 23 (2006) 1883-1912.

\bibitem{35} G. D. Barbosa and N. Pinto-Neto, Phys. Rev. D 70 (2004) 103512.

\bibitem{36} S. Zaim and L. Khodja, Phys. Scr. 81(2010) 055103.

\bibitem{37} N.~Mebarki, L.~Khodja, and S.~Zaim, EJTP 7,23(2010)181-196.

\bibitem{38} G. D. Barbosa and N. Pinto-Neto, Phys. Rev. D 70 (2004) 103512.

\bibitem{39} J. W. Moffat, Phys. Lett. B493 (2000) 142-148.

\bibitem{40} J. W. Moffat, Phys. Lett. B491 (2000) 345-352.

\bibitem{41} C. Acatrinei, Phys. Rev. D67 (2003) 045020.

\bibitem{42} D. V. Vassilevich, Nucl. Phys. B715 (2005) 695-712.

\bibitem{43} M. Rosenbaum, J. D. Vergara, L. R. Juarez, J. Phys. A A40
(2007) 10367-10382.

\bibitem{44} N.~Mebarki, S.~Zaim, L.~Khodja and H.~Aissaoui, Phys.\ Scripta 
\textbf{78} (2008) 045101.

\bibitem{45} P. Aschieri, B. Jurco, P. Schupp and J. Wess, Nucl. Phys. B651,
45 (2003).

\bibitem{46} P. Mukherjee and A. Saha, Phys. Rev. D74, 027702 (2006).

\bibitem{47} M.~Chaichian, M.~R.~Setare, A.~Tureanu and G.~Zet, JHEP \textbf{%
04} (2008) 064.

\bibitem{48} Slimane Zaim and Hadjar Rezki, Gravitation and Cosmology,
26(3):200--207, (2020).

\bibitem{49} A. Touati and S. Zaim, Chinese Physics C 46, 105101 (2022).

\bibitem{50} A. Touati and S. Zaim, Annals of Physics 455, 169394 (2023).

\bibitem{51} Rezki, H., Zaim, S. Theor Math Phys 219, 856--870 (2024).

\bibitem{52} Charles Schwartz, International Journal of Modern Physics
A.Vol. 31, No. 09, 1650041 (2016).

\bibitem{53} J. Schwinger, Phys, Rev. 82, 664 (1951).

\bibitem{54} J. Garriga, Phys.Rev.D 49 (1994) 6343-634.

\bibitem{55} Markus B. Fr\"{o}b et al JCAP 04 (2014)009.
\end{thebibliography}
\end{document}